\documentclass[aps,prl,twocolumn,groupedaddress]{revtex4}

\usepackage{graphics} 
\usepackage{graphicx}
\usepackage{dcolumn}
\usepackage{bm}

\setcitestyle{super}

\begin{document}

\title{Dynamics of Quantal Heating in Electron Systems with Discrete Spectra}

\author{Scott Dietrich}
\author{William Mayer}
\author{Sergey Vitkalov}
\email[Corresponding author: ]{vitkalov@sci.ccny.cuny.edu}
\affiliation{Physics Department, City College of the City University of New York, New York 10031, USA}
\author{A. A. Bykov}
\affiliation{A.V.Rzhanov Institute of Semiconductor Physics, Novosibirsk 630090, Russia}
\affiliation{Novosibirsk State University, Novosibirsk 630090, Russia}

\date{\today}

\begin{abstract} 

The temporal evolution of  quantal Joule heating of 2D electrons in GaAs quantum well placed in quantizing magnetic fields is studied using a difference frequency method. The method is based on measurements of the electron conductivity oscillating at the beat frequency $f=f_1-f_2$ between two microwaves applied to 2D system at  frequencies $f_1$ and $f_2$. The method provides  $direct$ access to the dynamical characteristics of the heating and yields  the inelastic scattering time $\tau_{in}$ of 2D electrons. The obtained  $\tau_{in}$ is  strongly temperature dependent, varying from 0.13 ns at 5.5K to 1 ns at 2.4K in magnetic field  $B$=0.333T.  When temperature $T$ exceeds the Landau level separation  the relaxation rate $1/\tau_{in}$ is  proportional to $T^2$, indicating the electron-electron interaction as the dominant mechanism limiting the quantal heating. At lower temperatures the rate tends to be proportional to $T^3$, indicating  considerable contribution from electron-phonon scattering.

\end{abstract}
 

\maketitle

Joule heating is a remarkable physical phenomenon which transforms electric energy into heat. Recently it was shown that the quantum properties of matter significantly affect   the heating \cite{dmitriev2005,zhang2009},  giving rise to a thermal stratification (quantization)  of the electron distribution  in energy space\cite{romero2008}. This effect, called quantal heating,  does not exist in classical electron systems. The most essential property of quantal heating is the conservation of the total number of quantum states participating in the electron transport and, thus,  the conservation of the overall broadening of the electron distribution\cite{zhang2009,romero2008}.   

In contrast to classical Joule heating,  quantal heating leads to outstanding nonlinear transport properties of highly mobile 2D electrons, driving  them  into  exotic nonlinear states in which voltage (current) does not depend on current \cite{bykov2007} (voltage)\cite{bykov2012}. Quantal heating  also provides significant contributions to nonlinear effects at high driving frequencies - a very important topic in contemporary research\cite{dmitriev2012}.   

The effects of  quantal heating on electron transport have been investigated  so far in the $dc$ domain. In the recent   investigations\cite{zhang2009}  the relaxation of the nonequilibrium electron distribution  was approximated by an inelastic scattering time $\tau_{in}$. The $\tau$-approximation  of the inelastic collision integral \cite{dmitriev2005} allows one to evaluate the inelastic relaxation rate and its temperature dependence from experiments at relatively high temperatures. The accuracy of the $\tau$-approximation in the case of separated Landau levels at low temperatures has not been investigated. Moreover,  previous studies of Joule heating executed on 2D systems with a low electron mobility indicate very different temperature dependence of the inelastic relaxation time in magnetic fields\cite{heating}. This raises a concern regarding the validity of the  relaxation time $\tau_{in}^{dc}$ obtained in the $dc$ domain using the procedures and approximations based on  the quantal heating approach\cite{dmitriev2005,zhang2009}. An alternative experimental method accessing the inelastic relaxation processes is certainly needed  to settle this important issue.

In this Letter we present an experimental method providing  $direct$ access to the temporal evolution of the electron transport affected by  quantal heating. The method yields the inelastic relaxation time $\tau_{in}$. At high temperatures the time is found to be in a good  agreement with the inelastic time obtained in $dc$ experiments\cite{zhang2009}. At low temperatures a disagreement between  these two times is observed.  Presented below results  significantly support  the existing theory\cite{dmitriev2005}.
   
The dynamics of quantal heating was studied in  a two-dimensional system of highly mobile electrons in Corbino geometry. The Corbino disc  with inner radius $r_1=$0.9 mm and outer radius $r_2=$1 mm was fabricated from a selectively doped  single GaAs quantum well sandwiched between AlAs/GaAs superlattice barriers.   The width of the well was 13 nm. The structure was grown by molecular beam epitaxy on a (100) GaAs substrate. AuGe eutectic  was used to provide electric contacts to the 2D electron gas.  The 2D electron system with electron density $n$=8$\cdot$10$^{15}$ m$^{-2}$ and mobility $\mu=$112 m$^{2}$/Vs at  T=4.8K was studied at different temperatures from 2.4K to 6K in magnetic fields up to 1T.  

\begin{figure}[t]
\includegraphics[width=3.4in]{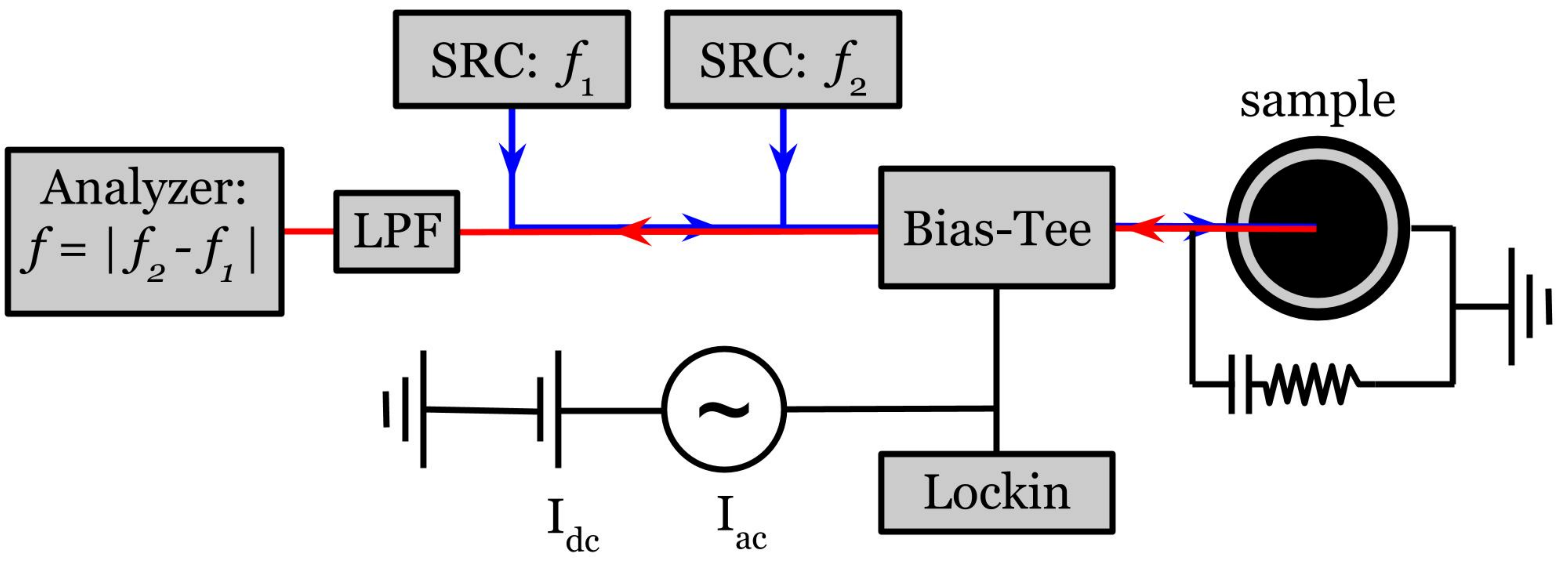}
\caption{(Color online) Experimental setup for the difference frequency method. Two microwave sources (SRC) at frequencies $f_1$ and $f_2$ are sent to the sample through a broadband directional couplers. The reflected signal is measured by the microwave  analyzer at the difference frequency of the two sources $f=f_2-f_1$.  Incorporated  $RC$ circuit ($R$=50 Ohm, $C$=47 pF)  provides broadband matching.  A low pass filter (LPF) blocks the high frequency signals  ($f_1, f_2$) from the analyzer. Bias current ($I_{dc}$) as well as low frequency lockin measurements ($I_{ac}$) are incorporated into the same universal measurement line through a bias-tee.}
\label{setup}
\end{figure}

Fig.\ref{setup} shows the experimental setup.  Two microwave sources supply the radiation  to the sample through a semi-rigid coax at two different frequencies ($f_1$, $f_2$).  The interference between these sources forms  microwave radiation with   amplitude modulation at the difference (beat) frequency $f=f_1-f_2$. The modulated microwave induces oscillations of Joule heating and, thus, the  sample conductance $\delta G_f$ at the frequency $f$. Application of a $dc$ voltage  $V_{dc}$ to the structure produces current oscillations $\delta I_f=\delta G_f V_{dc}$, which  propagate back to a microwave analyzer through the same coax. The analyzer detects the current oscillations  at  frequency $f$ ($f$-signal).  In addition the setup contains a bias-tee providing measurements in the $dc$ domain. These measurements are essential for a frequency calibration  of the microwave setup. 

In frequency experiments frequency $f_1=$ 8 GHz was fixed while frequency $f_2$ was scanned from 5.5 to  7.999 GHz. To take into account  variations of the microwave power $P_2$ delivered to the sample in the course of the frequency scan, a $dc$  measurement of the resistance variation induced  by the same applied microwave power $P_2$ is done. At a small applied power  the induced resistance variation is proportional to  $P_2$, thereby providing the power calibration.  A similar calibration is done for the receiver channel at frequency $f$ and is based on the reciprocal property of the microwave setup.    

\begin{figure}[t]
\includegraphics[width=3.4in]{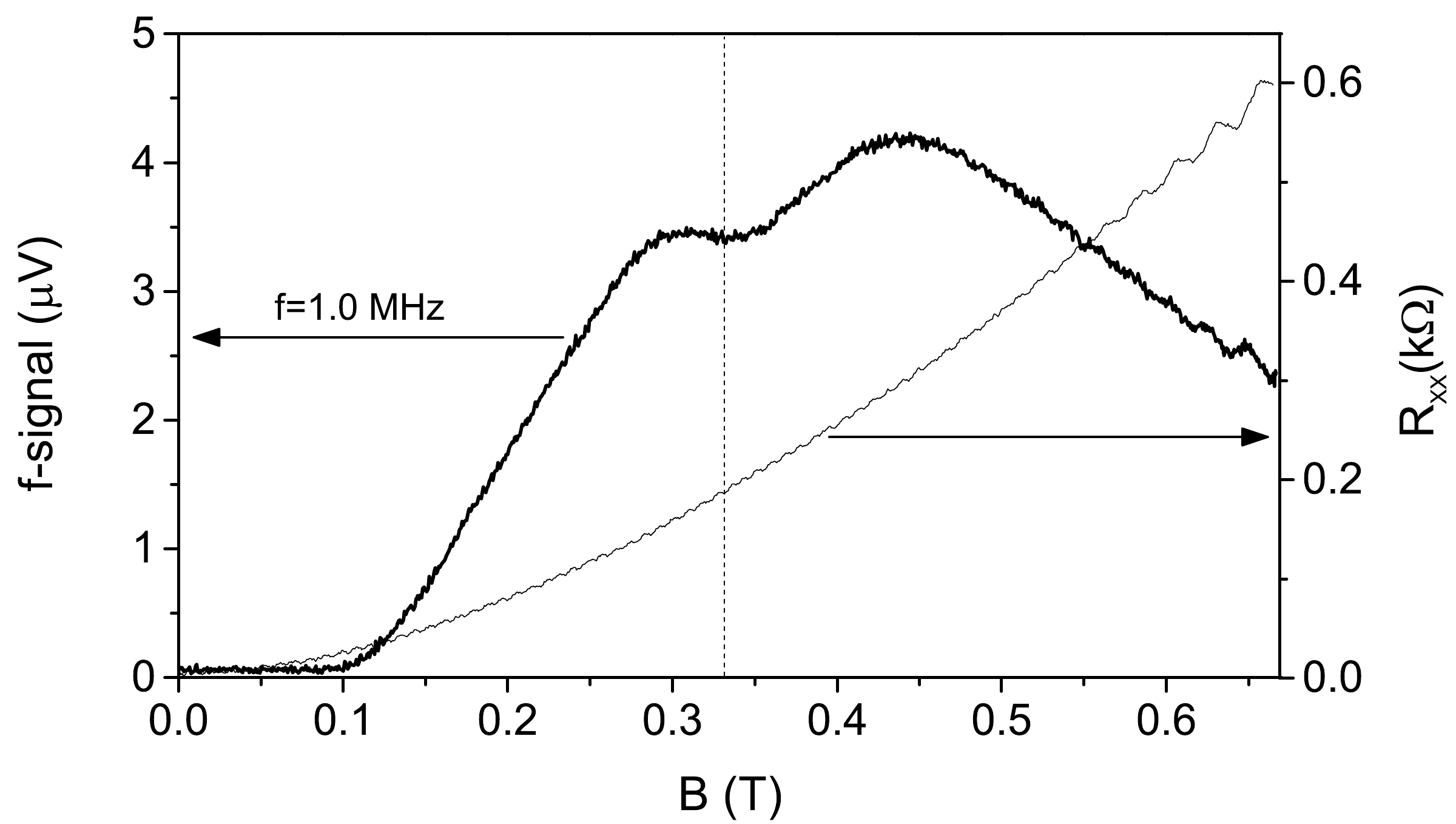}
\caption{ Magnetic field dependences of the sample resistance (right axis, thin line, no microwave and $dc$ bias applied) and microwave analyzer signal (left axis, thick line) at the difference frequency $f$ = 1.0 MHz   with MW sources at powers $P_1(8GHz)=-22$dBm and  $P_2(7.999GHz)=-19$dBm and with direct current $I_{dc}$=10 $\mu$A. The vertical dashed line indicates the magnetic field chosen for  study of the frequency dependence of the nonlinear response: $B$ = 0.333T.  $T$ = 4.8K.}
\label{Bfield}
\end{figure}

Figure \ref{Bfield} presents the magnetic field dependence of the resistance of the sample, $R$,   with neither $dc$ bias nor microwaves applied (thin solid line). As expected in the Corbino geometry the resistance shows the classical parabolic increase with the magnetic field $B$.   The thick solid line presents  the nonlinear response of the sample ($f$-signal) measured at difference frequency $f$=1 MHz. The nonlinear response is very weak at small magnetic fields $B<$0.1 T.  At these fields the Landau level separation $\hbar \omega_c$ is much smaller the levels width $\Gamma$ and both  the quantization of the electron spectrum and  quantal heating are absent\cite{zhang2009}.  Above 0.1 Tesla the Landau quantization occurs and quantal heating starts to grow, reaching maximums at about 0.3 and 0.45 T. At  higher magnetic fields the $f$-signal decreases due to a decrease of the cyclotron radius of electron orbits leading to significant reduction of the spatial and, thus, the spectral diffusions\cite{dmitriev2005}.    At T$=$4.8K and B$>$0.5T, Shubnikov de Haas (SdH) oscillations  are visible in both the resistance and the $f$-signal.  The frequency dependence of the $f$-signal was studied at magnetic field B=0.333 T corresponding to a maximum of the sample conductivity at low temperatures (not shown).  

\begin{figure}[t]
\includegraphics[width=3.4in]{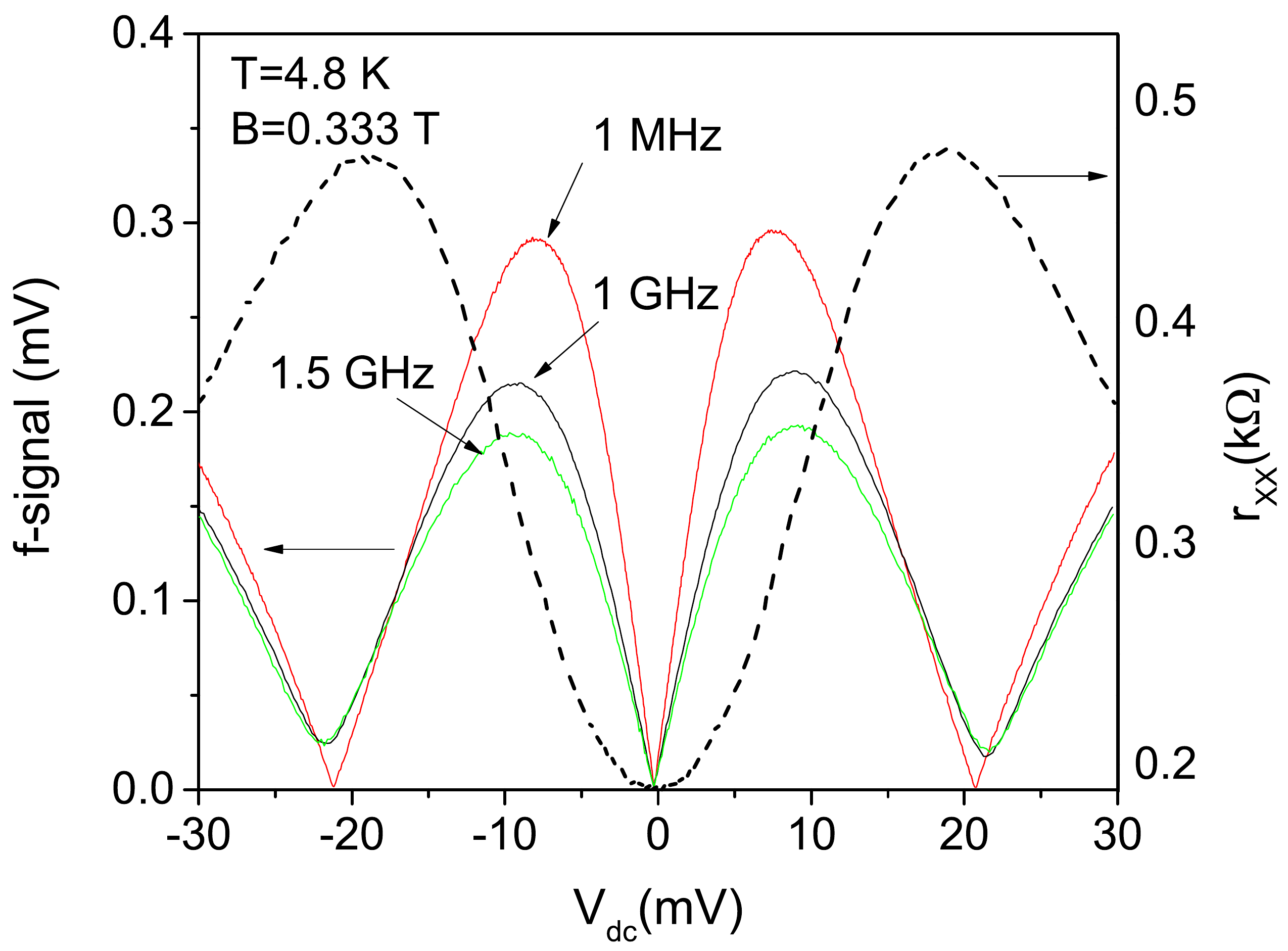}
\caption{Dependence of the $f$ signal on $dc$ voltage at different frequencies $f$ as labeled. Dashed line presents the $dc$ bias dependence of differential resistance $r_{xx}$ obtained in the $dc$-domain.  }
\label{dc}
\end{figure}

Figure \ref{dc} presents the dependence of the $f$-signal and differential resistance on the $dc$ bias at different frequencies $f$ as labeled. At small $dc$ biases $V_{dc}$ the $f$-signal is proportional to  $V_{dc}$ while the differential resistance $r_{xx} \sim V_{dc}^2$. These data agree with the relation $j=\sigma_0 E+\alpha E^3$ between the current density $j$ and the electric field $E$, which is  expected at small magnitude $E$.  Here $\sigma_0$ is the ohmic conductivity (linear response). In this perturbative regime the nonlinear current density $j_\omega$  at angular frequency $\omega=2\pi f$ (the $f$-signal) should be proportional to applied $dc$ ($E_0$) and $mw$ ($E_1, E_2$) electric fields: $j_\omega=\alpha E_0E_1E_2$.  The observed microwave power dependence (not shown) of the $f$-signal is in  complete agreement with the expected behavior at small microwave power.  At higher $dc$ biases the $f$-signal demonstrate an additional interesting features, which were, however, beyond the scope of the present work.

The frequency dependence of the nonlinear response can be understood from an analysis of the spectral diffusion equation for the electron distribution function $f(\epsilon)$\cite{zhang2009,dmitriev2005}:
\begin{equation}
-\frac{\partial f(\epsilon)}{\partial t}+E^2\frac{\sigma _D}{\nu (\epsilon)}\partial_{\epsilon}\left[\tilde{\nu} ^2(\epsilon) \partial _{\epsilon}f(\epsilon)\right]=\frac{f(\epsilon)-f_T(\epsilon)}{\tau_{in}}
\label{main}
\end{equation}
Here, $\sigma _D(B)$ is the Drude conductivity in a magnetic field $B$, $\tilde{\nu}=\nu/\nu_0$ is  ratio of the density of electron states (DOS) $\nu(\epsilon)$ to the DOS at zero magnetic field $\nu_0$ and $f_T$ represents the Fermi-Dirac distribution at a temperature $T$.  Below we consider the case of a low difference frequency $\omega=\omega_1-\omega_2 \ll \omega_1,\omega_2$ corresponding to the experiments. At small electric field $E(t)=E_0+E_1exp(i\omega_1t)+E_2exp(i\omega_2 t)$ the distribution function can be written as $f=f_T+\delta f_\omega$, where the oscillating distribution $\delta f_\omega \sim E_1E_2exp[i(\omega_1-\omega_2) t]$ is the leading contribution to the $f$-signal.  A substitution of this function into Eq.(\ref{main}) yields the following solution for the electron distribution oscillating at difference frequency $\omega$:
\begin{equation}
\delta f_\omega (\epsilon)=\frac{2E_1E_2exp(i\omega t)}{i\omega +1/\tau_{in}}\cdot \frac{\sigma _D}{\nu(\epsilon)}\partial_{\epsilon}\left[\tilde{\nu} ^2(\epsilon) \partial _{\epsilon}f_T(\epsilon)\right]
\label{distr}
\end{equation} 

The oscillating electron distribution results in oscillations of the electric current density at  frequency $\omega$ :
\begin{equation}
j_\omega=E_0\int  \sigma_\epsilon [-\partial_\epsilon (\delta f_\omega ) ] d\epsilon=
\frac{2E_0E_1E_2exp(i\omega t)}{i\omega +1/\tau_{in}}\cdot \Sigma 
\label{fcurr}
\end{equation}
Here $\sigma_\epsilon$ is the conductivity at energy $\epsilon$ and $\Sigma= -\sigma _D \int  \sigma_\epsilon \partial_\epsilon [ \partial_{\epsilon}(\tilde{\nu} ^2 \partial _{\epsilon}f_T)  /\nu ]d\epsilon$. Eq.(\ref{fcurr}) indicates that at high difference frequency $\omega \gg 1/\tau_{in}$ the $f$-signal is inversely proportional to  frequency. In this regime,  microwave radiation is "on" for a short time $\Delta t \sim 1/\omega$, which is not enough to considerably change  the electron distribution. 

\begin{figure}[t]
\includegraphics[width=3.1in]{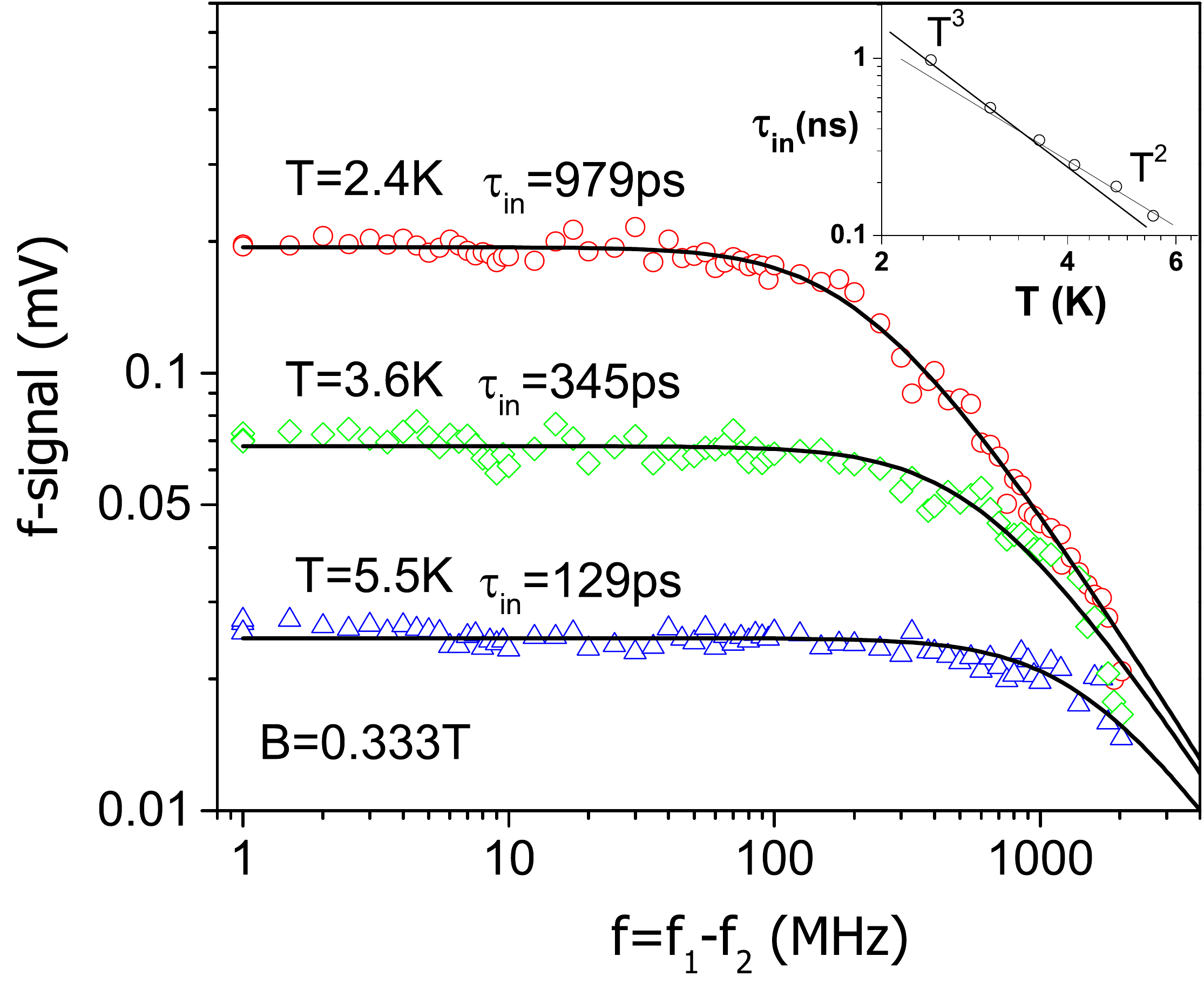}
\caption{Frequency dependence of the $f$-signal at different temperatures $T$ as labeled. Solid  lines presents the dependence obtained from Eq.(\ref{fcurr}) using $\tau_{in}$ as a fitting parameter. The insert presents the temperature dependence of the obtained inelastic scattering time. }
\label{fdep}
\end{figure}

Figure \ref{fdep} presents the frequency dependence of the $f$-signal at different temperatures as labeled. The observed $f$-signal is nearly frequency independent  at  low  frequencies and is inversely proportional to the frequency in the high frequency limit. The solid lines represent the frequency dependence expected from Eq.(\ref{fcurr}): $j_\omega=A/\vert 1+i \omega \tau_{in} \vert$ with  amplitude $A$  and time $\tau_{in}$ as  fitting parameters. The figure indicates a good agreement between the data and the frequency dependence described by  Eq.(\ref{fcurr}). The insert to the figure presents the temperature dependence of the inelastic scattering time $\tau_{in}$ obtained from the fit.  The high temperature behavior of the inelastic time is consistent with $T^2$ decrease indicating the dominant contribution of electron-electron interactions to the inelastic electron relaxation.  At low temperatures a deviation  from the $T^2$ behavior is found, indicating a suppression of the $e-e$ contribution.  The suppression is expected at low temperatures, when $kT< \hbar \omega_c$, where  $\omega_c$ is the cyclotron frequency \cite{zhang2009}.  

\begin{figure}[t]
\includegraphics[width=3.4in]{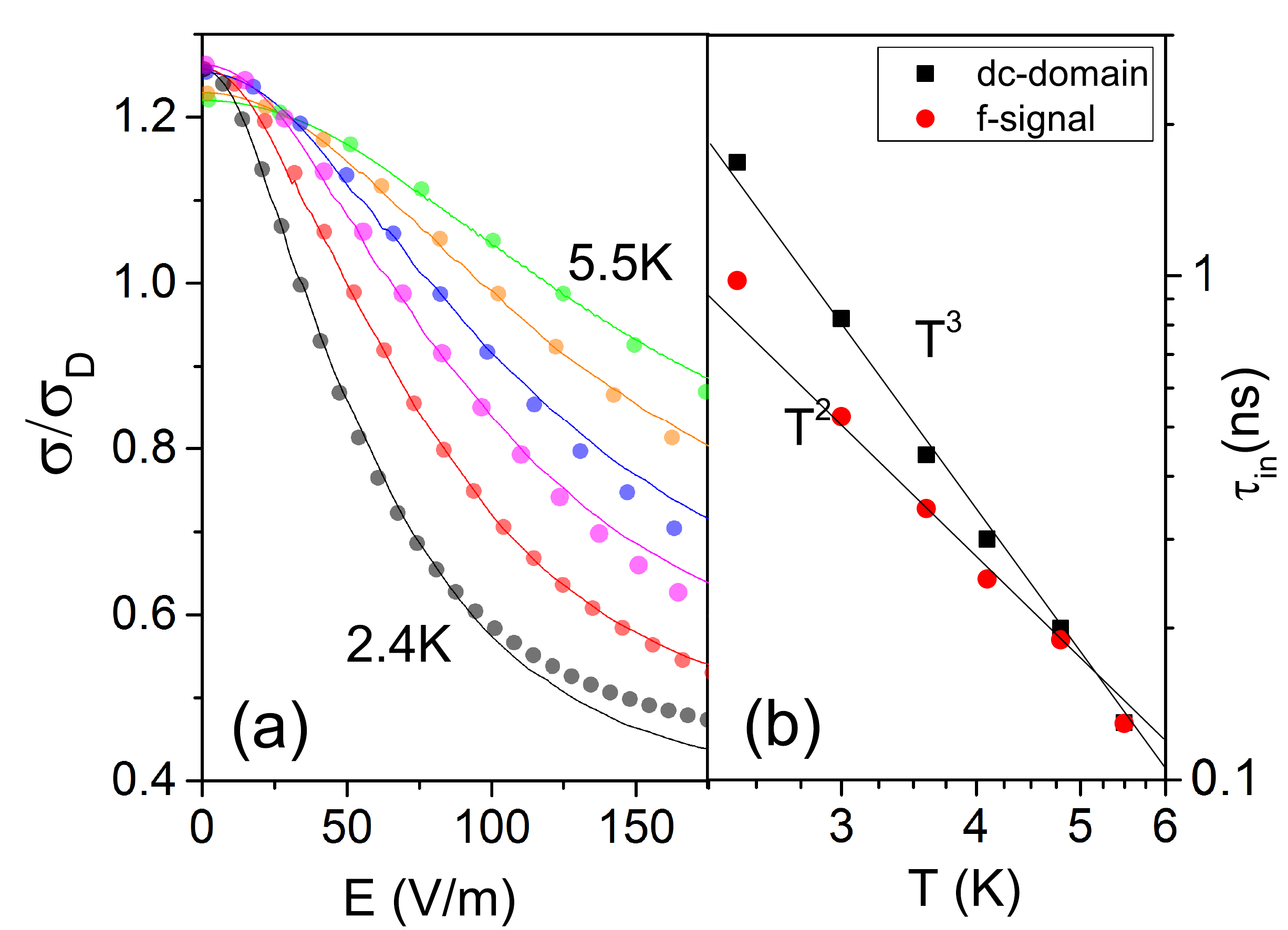}
\caption{ (a) Solid lines present the dependence of the normalized conductivity $\sigma/\sigma_D$ on electric field $E$ at different temperatures: 5.5, 4.8, 4.1, 3.6, 3.1 and 2.4K from top to bottom.  Symbols present  simulation of the nonlinear resistance based on the numerical solution of Eq.(\ref{main}) using Gaussian approximation for the electron density of states \cite{zhang2009}; (b) Temperature dependence of the inelastic scattering time obtained in high frequency experiments ($f$-signal shown in Fig.\ref{fdep}) and from the response  in the $dc$-domain shown in (a). $B$ = 0.333T. }
\label{dc_domain}
\end{figure}

It is important to compare the obtained inelastic time  $\tau_{in}$ with the time $\tau_{in}^{dc}$  obtained from  the nonlinear response in the $dc$ domain \cite{zhang2009}. Fig.\ref{dc_domain}(a) presents the dependence of the normalized conductivity of the sample $\sigma/\sigma_D$ on the applied electric field $E \approx V_{dc}/(r_2-r_1)$\cite{Efield}. In the Corbino geometry the Drude conductivity  $\sigma_D$ is obtained from a comparison of the magneto-conductivity of the sample $\sigma(B)$ with the expected classical behavior: $\sigma_D(B)\sim 1/(1+(\omega_c\tau_{tr})^2)$, where  $\tau_{tr}$ is the transport scattering time. Solid lines present experimental dependences obtained at different temperatures. Filled circles present results of numerical simulations of the nonlinear response based on the spectral diffusion equation (\ref{main}). The simulation uses a Gaussian approximation for the electron density of states \cite{zhang2009}. Comparison between the experiments and numerical simulations yields the inelastic relaxation time $\tau_{in}^{dc}$. 

Fig.\ref{dc_domain}(b) shows temperature dependencies of the time $\tau_{in}^{dc}$ and  the inelastic time obtained from the dynamics of the nonlinear response ($f$-signal).  At high temperatures both  times are close to each other.  At lower temperatures there is  a considerable difference between the two times suggesting that important parameters  responsible for the nonlinear response are not well understood or remain unknown.   The  $dc$-domain inelastic time $\tau_{in}^{dc}(T)$ follows $1/T^3$ decrease with the temperature $T$, while the time $\tau_{in}$, obtained from $f$-signal, is mostly proportional to $1/T^2$ with a tendency to $1/T^3$ at low temperatures.   

Previous $dc$ domain investigations  have shown that in the regime when only few Landau levels provide the electron transport ($kT < \hbar \omega_c$) the electron-electron scattering is ineffective for the relaxation of the overheated electrons  since the scattering conserves the total electron energy\cite{zhang2009}.  In this regime the experiments show the $1/T^3$ temperature dependence of the time $\tau_{in}^{dc}(T)$  attributed to contributions of the electron-phonon interaction to the electron energy relaxation\cite{zhang2009}. At small magnetic fields ($kT \gg \hbar \omega_c$) the inelastic time $\tau_{in}^{dc}(T)$ was found to follow $1/T^2$ dependence attributed to electron-electron scattering\cite{zhang2009} in accord with the theory\cite{dmitriev2005}.   The transition between two regimes is poorly understood.  The presented results have been obtained in the transitional  regime ($kT \sim \hbar \omega_c$)\cite{f_range}. The discrepancy between $\tau_{in}(T)$  obtained in the  high frequency  and  the $dc$ domains  suggests  the complexity of actual inelastic processes in this 2D system.
         
Furthermore, another mechanism of nonlinear transport, which is comparable with the effects of quantal heating, has been recently identified  in the regime of SdH oscillations \cite{dietrich2012,dietrich2013}. This mechanism was related to $dc$ bias induced electron spatial redistribution which may  have very different (most likely very slow) relaxation  and, thus, may not have been captured by the $f$-signal in the studied frequency range\cite{redistr}. 

Shown in Fig.\ref{dc_domain} (b),   the discrepancy between the  two times  may also  be related  to  assumptions and approximations used to extract the inelastic scattering rate in the $dc$ domain. In particular the expression for conductivity used in previous investigations \cite{zhang2009}   $\sigma(\epsilon)=\sigma_D\tilde{\nu} ^2$ may not be quite adequate for separated Landau levels. In contrast to the $dc$ domain result,  the frequency dependence of the $f$-signal is a direct measurement and is not  subject to assumptions and/or approximations. Thus the dynamical measurements provide more accurate and reliable information on the  inelastic relaxation of 2D electrons placed in quantizing magnetic fields. 

In summary  the dynamics of the strongly nonlinear response of 2D highly mobile electrons placed in quantizing magnetic fields is observed.  The obtained results  indicate  the dominant conribution of quantal heating to the response and significantly support  the existing theory\cite{dmitriev2005}.  The presented method provides the direct measurement of  the inelastic relaxation in  2D electron systems.

This work was supported by the National Science Foundation (DMR 1104503),  the Russian Foundation for Basic Research (project no.14-02-01158) and  the Ministry of Education and Science of the Russian Federation.


\begin{thebibliography} {8}
\bibitem{Efield} In the Corbino geometry the electric field $E(r) \sim ln(r)$ varies with the  radius $r$. For Corbino disc with a narrow conducting channel $ r_2-r_1 \ll r_1, r_2$ the field variation is small and the  electric field is approximated by an average value $E\approx V_{dc}/(r_2-r_1)$.
\bibitem{redistr} In a magnetic field corresponding to a maximum of the conductivity (B=0.333T) and at small $dc$ biases  the redistribution mechanism and the quantal heating work together enhancing the total nonlinearity and narrowing the conductivity drop shown in Fig.\ref{dc_domain}(a). It can make the extracted $\tau_{in}^{dc}$ to be longer.
\end{thebibliography}
\end{document}